\documentclass[12pt]{iopart}

\usepackage{iopams}
\usepackage{graphicx}
\usepackage{mathrsfs}
\usepackage{color}
\usepackage[table]{xcolor}
\usepackage{hyperref}
\hypersetup{colorlinks, citecolor=black, filecolor=black, linkcolor=black, urlcolor=black}

\expandafter\let\csname equation*\endcsname\relax
\expandafter\let\csname endequation*\endcsname\relax
\usepackage{amsmath}

\usepackage{dcolumn}
\usepackage{bm}

\definecolor{grayish}{RGB}{230,230,230}

\newcommand{\refEq}[1] {(\ref{#1})}
\newcommand{\superscript}[1]{\ensuremath{^{\textrm{#1}}}}

\newcommand{\Sin}[1]{\ensuremath{\sin \left( #1 \right)}}
\newcommand{\Cos}[1]{\ensuremath{\cos \left( #1 \right)}}

\newcommand{\Tan}[1]{\ensuremath{\tan \left( #1 \right)}}

\newcommand{\Ln}[1]{\ensuremath{\ln \left( #1 \right)}}

\newcommand{\Nabla}{\ensuremath{\vec{\nabla}}}
\newcommand{\romanNum}[1]{\uppercase\expandafter{\romannumeral#1}}

\begin{document}

\title{Optimized up-down asymmetry to drive fast intrinsic rotation in tokamaks}

\author{Justin Ball\superscript{1,2,3}, Felix I. Parra\superscript{2,3}, Matt Landreman\superscript{4}, and Michael Barnes\superscript{2,3}}

\address{\superscript{1} Swiss Plasma Center, \'{E}cole Polytechnique F\'{e}d\'{e}rale de Lausanne, 1015 Lausanne, Switzerland}
\address{\superscript{2} Rudolf Peierls Centre for Theoretical Physics, University of Oxford, Oxford OX1 3NP, United Kingdom}
\address{\superscript{3} Culham Centre for Fusion Energy, Culham Science Centre, Abingdon OX14 3DB, United Kingdom}
\address{\superscript{4} Institute for Research in Electronics and Applied Physics, University of Maryland, College Park, Maryland 20742, USA}
\ead{Justin.Ball@physics.ox.ac.uk}

\begin{abstract}

Breaking the up-down symmetry of the tokamak poloidal cross-section can significantly increase the spontaneous rotation due to turbulent momentum transport. In this work, we optimize the shape of flux surfaces with both tilted elongation and tilted triangularity in order to maximize this drive of intrinsic rotation. Nonlinear gyrokinetic simulations demonstrate that adding optimally-tilted triangularity can double the momentum transport of a tilted elliptical shape. This work indicates that tilting the elongation and triangularity in an ITER-like device can reduce the energy transport and drive intrinsic rotation with an Alfv\'{e}n Mach number of roughly $1\%$. This rotation is four times larger than the rotation expected in ITER and is approximately what is needed to stabilize MHD instabilities. It is shown that this optimal shape can be created using the shaping coils of several present-day experiments.

\end{abstract}



\section{Introduction}
\label{sec:introduction}

Bulk toroidal rotation is usually beneficial for plasma performance in tokamaks. Rotation with an Alfv\'{e}n Mach number of $M_{A} \approx 1\%$ is able to stabilize dangerous magnetohydrodynamic (MHD) modes \cite{LiuITERrwmStabilization2004, GarofaloExpRWMstabilizationD3D2002, ReimerdesRWMmachineComp2006}. This corresponds to a Mach number of roughly $M_{S} \equiv \sqrt{2 T_{i}/m_{i}} \approx 5 \%$ in ITER \cite{AymarITERSummary2001}, where $T_{i}$ is the ion temperature and $m_{i}$ is the ion mass. Moreover, higher levels of rotation can combat turbulence \cite{RitzRotShearTurbSuppression1990, BurrellShearTurbStabilization1997} (though extremely high rotation can actually drive turbulence through the parallel velocity gradient instability \cite{CattoPVG1973,MattorPVG1988,NewtonPVG2010}). Unfortunately, the mechanisms that drive toroidal rotation in existing experiments do not appear to scale well to future high-performance devices like ITER (i.e. larger devices with stronger magnetic fields).

Rotation is commonly driven by pushing the plasma using external injection of momentum. This is often done with beams of neutral particles, which enable existing experiments to achieve toroidal rotation with typical values of $M_{A} \approx 3 \%$ (or $M_{S} \approx 15\%$) \cite{GarofaloExpRWMstabilizationD3D2002}. However, due to ITER's larger size, external injection is only expected to drive rotation with $M_{A} \approx 0.3 \%$ (or $M_{S} \approx 1.5\%$) \cite{LiuITERrwmStabilization2004}, significantly less than what is needed for MHD stabilization.

\begin{figure}
	\centering
	\includegraphics[width=\columnwidth]{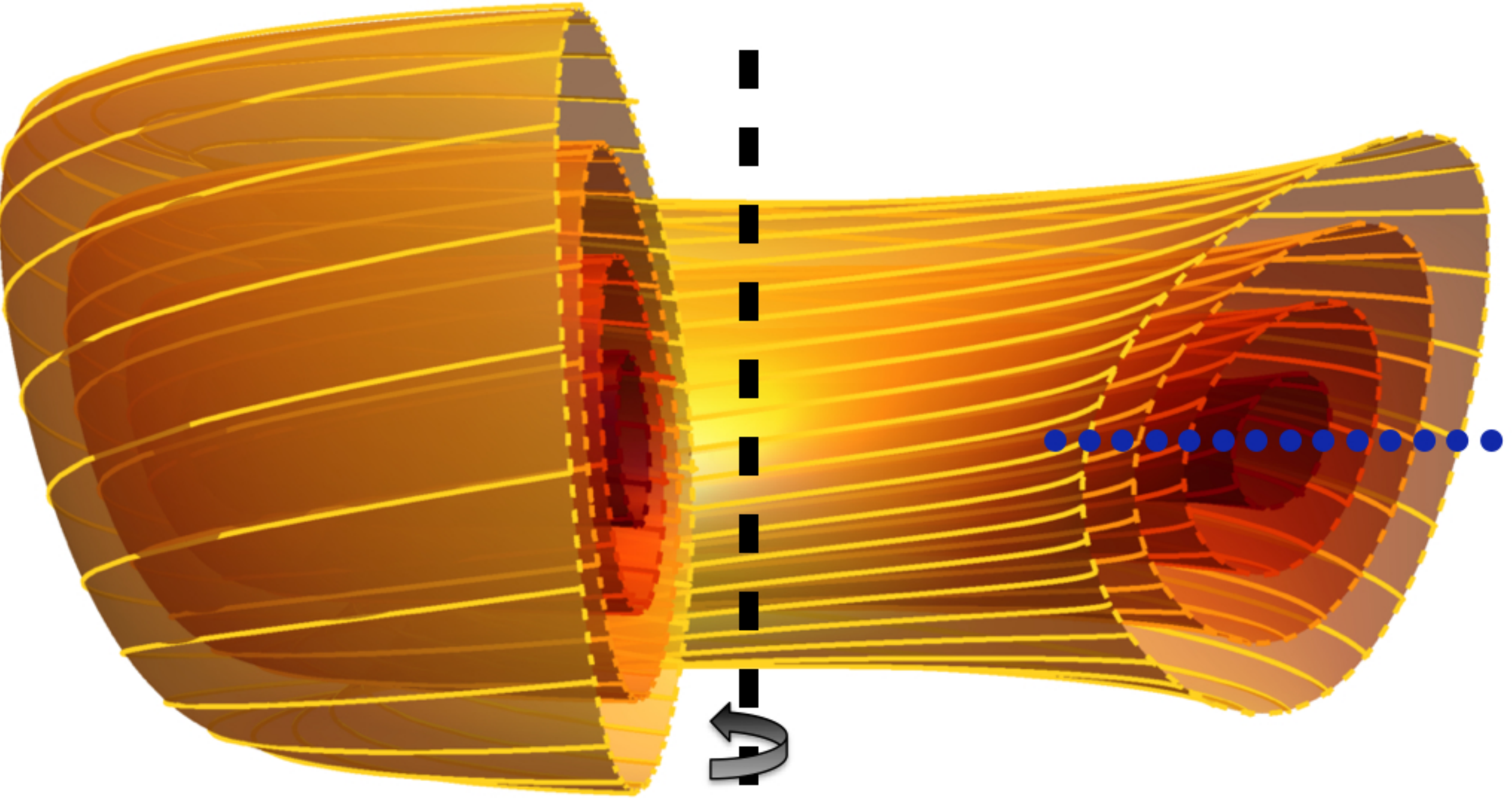}
		
	\caption{A cutaway view of the flux surfaces in the ``optimal'' magnetic geometry (i.e. the equilibrium that will be shown to maximize intrinsic toroidal rotation generated by up-down asymmetry), shown with the midplane (blue, dotted) and axis of toroidal symmetry (black, dashed).}
	\label{fig:cutawayTokamak}
\end{figure}

An attractive alternative is ``intrinsic'' rotation, or rotation spontaneously generated by the turbulent transport of momentum. Turbulence can move momentum around the device, just like it moves particles and energy. This enables the nested magnetic flux surfaces to push off each another, as well as the surrounding vacuum vessel and external magnets. Unfortunately, the intrinsic rotation in existing experiments is observed to be fairly modest, typically with $M_{A} \lesssim 1 \%$ (or $M_{S} \lesssim 5\%$) \cite{RiceExpIntrinsicRotMeas2007}. This has been explained by noting a particular symmetry \cite{PeetersMomTransSym2005, ParraUpDownSym2011, SugamaUpDownSym2011} of the gyrokinetic model, a theoretical model that is thought to accurately describe turbulence in the core of tokamaks. This symmetry constrains the intrinsic rotation to be small in $\rho_{\ast} \equiv \rho_{i} / a \ll 1$, where $\rho_{i}$ is the ion gyroradius and $a$ is the tokamak minor radius (strictly speaking, intrinsic rotation is small in the ion {\it poloidal} gyroradius divided by the minor radius \cite{ParraIntrinsicRotTheory2015}). However, it is broken if the magnetic equilibrium is up-down asymmetric (i.e. is not mirror symmetric about the midplane as shown in figure \ref{fig:cutawayTokamak}). Subsequent numerical \cite{CamenenPRLSim2009, BallMomUpDownAsym2014} and experimental \cite{CamenenPRLExp2010} work using flux surfaces with tilted elliptical shapes suggests that up-down asymmetry is a feasible method to generate the present experimentally-measured rotation levels in future larger devices (which have smaller values of $\rho_{\ast}$).

Recently, a series of analytic arguments have been formulated concerning the ability of different up-down asymmetric flux surface shapes to drive intrinsic rotation. MHD analysis has shown that externally-applied shaping with a large poloidal mode number $m \gg 1$ does not penetrate effectively throughout the plasma, but only shapes an exponentially thin layer at the edge \cite{BallDoctoralThesis2016}. Furthermore, another symmetry of the gyrokinetic model, a poloidal tilting symmetry, has been demonstrated in the limit of high-order flux surface shaping \cite{BallMirrorSymArg2016}. The tilting symmetry implies that mirror symmetric flux surfaces (i.e. flux surfaces that have mirror symmetry about at least one line in the poloidal plane) drive momentum transport that is exponentially small in $m \gg 1$. There are two ways to circumvent this restriction. First, we can break tilting symmetry using non-mirror symmetric flux surface shapes \cite{BallMomFluxScaling2016,BallEnvelopes2017} (i.e. shapes that do not have mirror symmetry about any line in the poloidal plane). This allows rotation to be driven by the direct interaction between different shaping modes (irrespective of toroidicity). This mechanism is expected to dominate in the limit of cylindrical devices. Second, we can beat two different shaping effects together in order to create an envelope that gives the flux surface shape low-order variation. Reference \cite{BallMomFluxScaling2016} indicates that, as long as the envelope is up-down asymmetric, this strategy can drive rotation that is only polynomially small in $m \gg 1$. This mechanism is expected to dominate in the limit of high-order shaping effects.

Taken together all these arguments motivate up-down asymmetric flux surface shapes with the lowest possible poloidal mode numbers. Additionally, it is important to explore the effects of non-mirror symmetry and up-down asymmetric envelopes, as they may enhance the rotation drive. In this work, we will consider flux surfaces with both elongation and triangularity. We will allow the two shaping effects to be tilted independently in order to break mirror symmetry and create up-down asymmetric envelopes. These two modes, $m = 2$ and $m = 3$, are the lowest order modes that can be created by external shaping coils. The $m = 1$ mode (i.e. the Shafranov shift) is not directly controlled by external magnets and has already been considered in reference \cite{BallShafranovShift2016}. By varying these two tilt angles, we will optimize the geometry to drive the fastest intrinsic rotation and look for geometries that directly improve the energy confinement time (irrespective of rotation).

\section{Magnetic equilibrium}
\label{sec:magneticEquil}

Before we can simulate turbulence, we must first specify the magnetic equilibrium. Due to axisymmetry, the general form of the magnetic field in a tokamak is given by
\begin{align}
   \vec{B} = I \left( \psi \right) \Nabla \zeta + \Nabla \zeta \times \Nabla \psi , \label{eq:magField} .
\end{align}
where $\zeta$ is the toroidal angle and $I \left( \psi \right)$ (which is closely related to the poloidal plasma current) controls the toroidal field strength. The magnetic flux surfaces (see figure \ref{fig:cutawayTokamak}) are contours of the stream function $\psi$ (which is the poloidal magnetic flux divided by $2 \pi$). We will motivate experimentally-practical flux surface shapes using solutions to the Grad-Shafranov equation
\begin{align}
   R^{2} \Nabla \cdot \left( \frac{\Nabla \psi}{R^{2}} \right) = - \mu_{0} R^{2} \frac{d p}{d \psi} - I \frac{d I}{d \psi} , \label{eq:gradShafEq}
\end{align}
which governs the tokamak equilibrium \cite{ShafranovGradShafranovEq1966}. Here $R$ is the major radial coordinate, $\mu_{0}$ is the permeability of free space, and $p$ is the plasma pressure. From equation \refEq{eq:magField} and Ampere's law, one can show that
\begin{align}
  - \mu_{0} R^{2} \frac{d p}{d \psi} - I \frac{d I}{d \psi} = \mu_{0} j_{\zeta} R ,
\end{align}
where $j_{\zeta}$ is the toroidal plasma current density. To find a simple and realistic solution, we assume $j_{\zeta}$ is uniform and expand in the limit of large aspect ratio (given the typical orderings for a low $\beta$, ohmically-heated tokamak \cite{FreidbergIdealMHD1987pg126}). In this context, equation \refEq{eq:gradShafEq} is solved by
\begin{align}
   \psi_{N} \left( r_{N}, \theta \right) &= r_{N}^{2} + \sum _{m=2}^{\infty} C_{N m} r_{N}^{m} \Cos{m \left( \theta + \theta_{m} \right)} , \label{eq:gradShafSol}
\end{align}
where we have normalized all lengths to the tokamak minor radius $a$, all magnetic fields to $\mu_{0} j_{\zeta} R_{0} / 4$, and indicated these normalized quantities by the subscript $N$. Here $r$ is the distance from the magnetic axis (the magnetic axis is the line enclosed by all flux surfaces), $\theta$ is the usual cylindrical poloidal angle, $C_{m}$ controls the strength of each shaping mode, $\theta_{m}$ is the mode tilt angle, and $R_{0}$ is the major radial location of the magnetic axis. Instead of using $\psi_{N}$ to label flux surfaces, we choose to use the flux surface label $\rho$, defined by
\begin{align}
   \psi_{N} \left( \rho \right) =  \rho^{2} + \sum_{m=2}^{\infty} C_{N m} \rho^{m} . \label{eq:rhoDef}
\end{align}
This substitution is made for convenience, as $\rho$ keeps the volume enclosed by the flux surfaces roughly constant as the tilt angles are changed. Specifically, $\rho$ corresponds to the value of $r_{N}$ at the outboard midplane if all shaping effects were untilted. To determine the shape of the flux surfaces, we must invert equation \refEq{eq:gradShafSol} in order to find $r_{N} \left( \rho, \theta \right)$. We will do this in two ways.

\begin{figure}
	\centering
	\includegraphics[width=0.4\columnwidth]{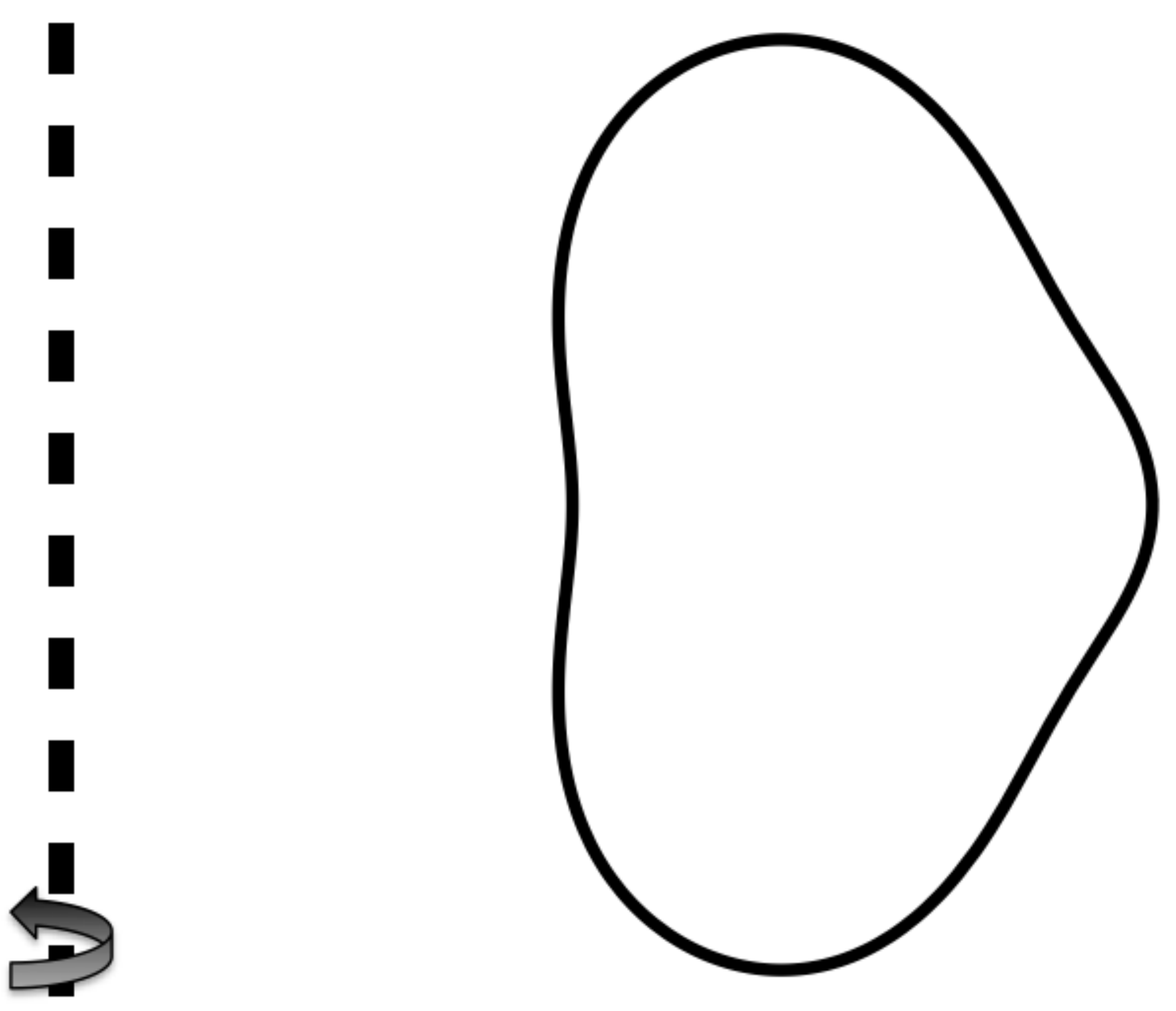}
		
	\caption{An example flux surface shape (solid) specified by the less-realistic parameterization of equation \refEq{eq:fluxSurfSpecGenElong}, along with the axis of toroidal symmetry (dashed). Note the slight concave region on the inboard side of the surface.}
	\label{fig:genElongShapeExample}
\end{figure}

The first shape parameterization is an approximate solution given by
\begin{align}
   r_{N} \left( \rho, \theta \right) = \sqrt{\psi_{N} \left( \rho \right)} \left( 1 + \sum_{m=2}^{\infty} \left( \frac{1}{\sqrt{1 + C_{N m} \rho^{m-2} \Cos{m \left( \theta + \theta_{m} \right)}}} - 1 \right) \right) . \label{eq:fluxSurfSpecGenElong}
\end{align}
This flux surface parameterization is constructed to be consistent with equation \refEq{eq:gradShafSol} in the weak shaping limit (i.e. $C_{N m} \ll 1$), to have an $m=2$ mode that exactly corresponds to an ellipse, and to include all shaping modes. However, because we are interested in the lowest order shaping effects, in this work we will only need the $m=2$ and $m=3$ terms. This parameterization is somewhat unrealistic because we will use fairly strong, ITER-like shaping. Since we are no longer respecting the weak shaping limit, equation \refEq{eq:fluxSurfSpecGenElong} can lead to concave regions in the flux surface shape (see figure \ref{fig:genElongShapeExample}), which are difficult to achieve experimentally. For this reason, we will only use the parameterization of equation \refEq{eq:fluxSurfSpecGenElong} to test the sensitivity of our results to the exact shape of the flux surface.

The second shape parameterization takes advantage of the fact that, when we only include the $m=2$ and $m=3$ modes, equation \refEq{eq:gradShafSol} becomes a cubic in $r$. This can be inverted exactly to find a more complicated parameterization for the shape of each flux surface:
\begin{align}
   r_{N} \left( \rho, \theta \right) &= \frac{1 + C_{N 2} \Cos{2 \left( \theta + \theta_{2} \right)}}{3 C_{N 3} \Cos{3 \left( \theta + \theta_{3} \right)}} \left( \Cos{\frac{\vartheta \left( \rho, \theta \right)}{3}} + \sqrt{3} \Sin{\frac{\vartheta \left( \rho, \theta \right)}{3}} - 1 \right) , \label{eq:fluxSurfSpecGlobal}
\end{align}
where
\begin{align}
   \Tan{\vartheta \left( \rho, \theta \right)} \equiv ~ & 3 C_{N 3} \sqrt{3 \psi_{N} \left( \rho \right)} \Cos{3 \left( \theta + \theta_{3} \right)} \times \label{eq:rAng} \\
   &\frac{\sqrt{4 \left( 1 + C_{N 2} \Cos{2 \left( \theta + \theta_{2} \right)} \right)^{3} - 27 C_{N 3}^{2} \psi_{N} \left( \rho \right) \cos^2 \left( 3 \left( \theta + \theta_{3} \right) \right)}}{2 \left( 1 + C_{N 2} \Cos{2 \left( \theta + \theta_{2} \right)} \right)^{3} - 27 C_{N 3}^{2} \psi_{N} \left( \rho \right) \cos^{2} \left( 3 \left( \theta + \theta_{3} \right) \right)} . \nonumber
\end{align}
We expect this second geometry specification to produce experimentally-practical flux surface shapes to lowest order in the large aspect ratio expansion. To show why, we first note that equation \refEq{eq:gradShafSol} is a solution to the Grad-Shafranov equation in both the vacuum and plasma regions. The only difference is that $j_{\zeta} = 0$ in vacuum, causing the inhomogeneous $r_{N}^{2}$ term to disappear. Studying equation \refEq{eq:gradShafSol}, we see that the low $m$ terms dominate in the $r_{N} \rightarrow 0$ limit. In fact, the $r_{N}^{m}$ dependence implies that high $m$ shaping modes are exponentially less effective at maintaining their effect over large distances in vacuum and plasma \cite{BallDoctoralThesis2016}. Hence, if the shaping coils are very far from the plasma (as would be preferable in a reactor), the only feasible shapes are those composed of the lowest-order modes in equation \refEq{eq:gradShafSol}. Equation \refEq{eq:fluxSurfSpecGlobal} is precisely that.

\begin{figure}
    \begin{flushleft}
        (a) \hspace{0.47\textwidth} (b)
    \end{flushleft}
    \begin{center}
        \includegraphics[width=\columnwidth]{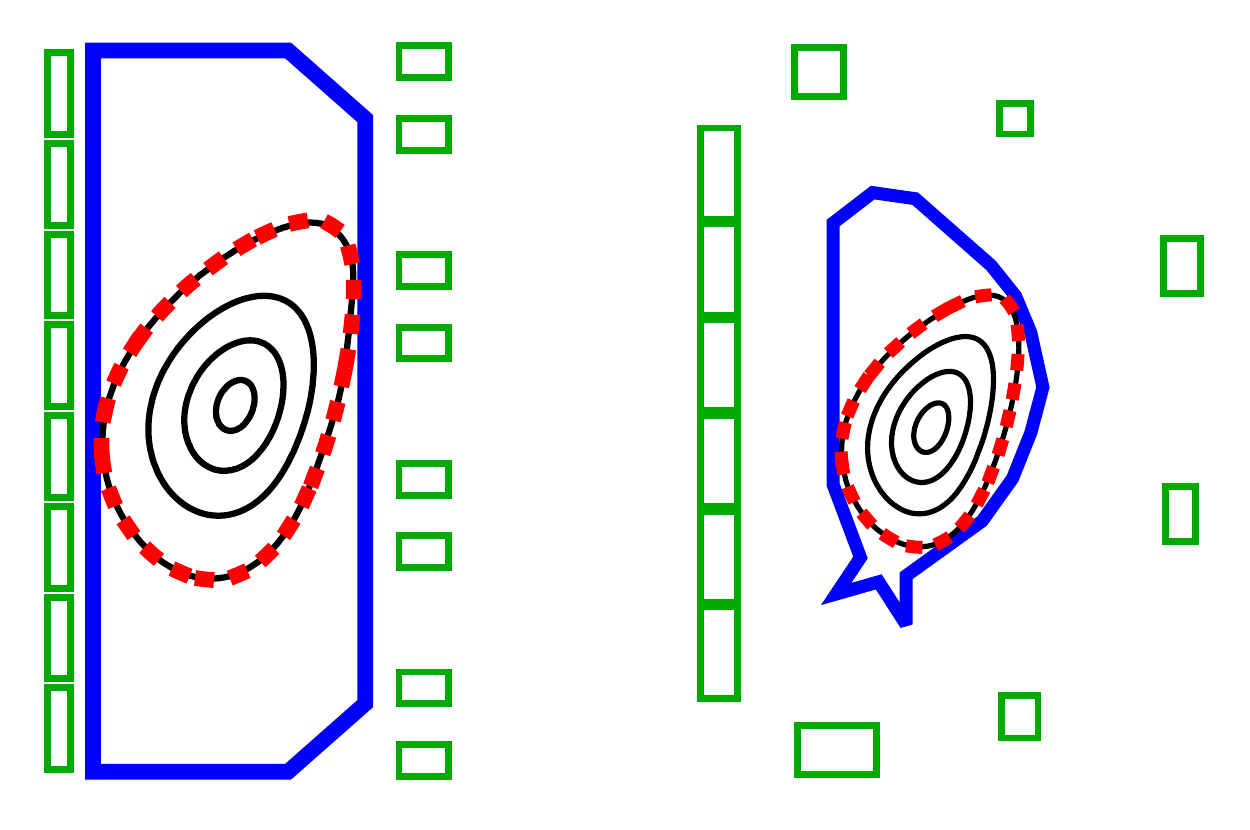}
    \end{center}
	\caption{Global free-boundary MHD equilibria (black, solid) created by the (a) TCV or (b) ITER poloidal shaping coils (green, rectangles) compared to the target ``optimal'' boundary shape (red, dashed, thick) along with the first wall (blue, solid, thick).}
	\label{fig:coilCurrents}
\end{figure}

To investigate experimental feasibility, we took the TCV \cite{HofmannTCVOverview1994} and ITER \cite{AymarITERSummary2001} coil sets and calculated the coil currents that most closely create our ``optimal'' flux surface shape. This ``optimal'' shape will be shown to maximize intrinsic rotation from up-down asymmetry, is parameterized by equation \refEq{eq:fluxSurfSpecGlobal}, and is shown in figure \ref{fig:cutawayTokamak}. To calculate the necessary coil currents, we must solve the inverse free-boundary equilibrium problem, which will be done in two ways. First, we used the FBT code \cite{HofmannFBT1988}, a free boundary MHD equilibrium code used to design magnetic geometries for the TCV experiment. This produced the TCV configuration shown in figure \ref{fig:coilCurrents}(a), which closely matches our ``optimal'' shape and respects all TCV coil limits for values of the plasma current up to 400 kA. Second, we employed VMEC calculations \cite{HirshmanVMEC1983} and performed a least-squares minimization to best approximate the target flux surface shape. This method is explained in the discussion of figure 9 in reference \cite{LandremanShapeCreation2016}. Our results (shown in figure \ref{fig:coilCurrents}(b)) demonstrate that the ITER coil set is capable of matching the ``optimal'' shape. However, we see that the mismatch between the plasma and vacuum vessel shapes leads to a reduction in the total plasma volume (compared to a nominal ``D''-shaped plasma).

\section{Intrinsic rotation driven by turbulence}
\label{sec:intrinsicRotation}

To model turbulent transport we will use gyrokinetics \cite{LeeGenFreqGyro1983, LeeParticleSimGyro1983, DubinHamiltonianGyro1983, HahmGyrokinetics1988, SugamaGyroTransport1996, SugamaHighFlowGyro1998, BrizardGyroFoundations2007, ParraGyrokineticLimitations2008, ParraLagrangianGyro2011, AbelGyrokineticsDeriv2012} because experimental measurements \cite{McKeeTurbulenceScale2001} indicate that it accurately treats turbulence in the core of tokamaks. Gyrokinetics is a fully-kinetic description based on an expansion of the Fokker-Planck and Maxwell's equations in $\rho_{\ast} \ll 1$. It specifically investigates behavior much slower than the ion gyrofrequency, but still allows the size of the turbulent eddies perpendicular to the magnetic field to be comparable to the gyroradius. In this regime, we can gyroaverage over the fast particle gyromotion. This removes one dimension of velocity space as well as the gyrofrequency timescale, which makes the model computationally tractable.

To further simplify, we will assume the turbulent fluctuations are electrostatic, implying that the magnetic field is not perturbed by the turbulent fluctuations. This is justified when the ratio of the thermal pressure to the magnetic pressure is small (i.e. the plasma beta is $\beta \ll 1$), which is fairly well satisfied in most experiments. We also assume that the turbulent fluctuations, which have perpendicular sizes similar to the ion gyroradius, are much smaller than the equilibrium gradients in density and temperature. Additionally, we are interested in mechanisms that cause a stationary plasma to begin rotating, so we will take the rotation and rotation shear to be zero (i.e. $\Omega_{\zeta} = d \Omega_{\zeta} / d \psi = 0$). Next, we will neglect particle collisions and instead use hyper-viscosity to provide enhanced numerical diffusion. This is done to reduce computational cost and because previous work suggested that it has little effect on our results \cite{BallMastersThesis2013}. Lastly, we will separate the perturbation in the distribution function due to turbulence $\delta f_{s}$ and assume it to be small by one order in $\rho_{\ast}$ compared to the background distribution function $F_{M s}$ (which we assume to be Maxwellian). Here the subscript $s$ indicates either the ion or electron species.

Given these assumptions, the gyrokinetic equation is given by \cite{BarnesCriticalBalance2011}
\begin{align}
   \frac{\partial \left\langle \delta f_{s} \right\rangle_{\varphi}}{\partial t} + \left( v_{||} \hat{b} + \vec{v}_{B} + \left\langle \vec{v}_{E} \right\rangle_{\varphi} \right) \cdot \Nabla \left( \left\langle \delta f_{s} \right\rangle_{\varphi} + \frac{Z_{s} e \left\langle \phi \right\rangle_{\varphi}}{T_{s}} F_{M s} \right) = - \left\langle \vec{v}_{E} \right\rangle_{\varphi} \cdot \Nabla F_{M s} .
\end{align}
This is solved together with the gyroaveraged, perturbed quasineutrality equation (i.e. Gauss's law)
\begin{align}
   \sum_{s} Z_{s} e \int_{- \infty}^{\infty} d v_{||} \int_{0}^{\infty} d \mu \oint d \varphi ~ \delta f_{s} = 0
\end{align}
for the perturbed distribution function and the electrostatic potential $\phi$. Here $t$ is the time, $\left\langle \ldots \right\rangle_{\varphi}$ is the gyroaverage at fixed guiding center position, $v_{||}$ is the component of the velocity parallel to the magnetic field unit vector $\hat{b}$, $\vec{v}_{B}$ is the magnetic drift velocity, $\vec{v}_{E}$ is the $\vec{E} \times \vec{B}$ drift velocity, $Z_{s}$ is the particle charge number, $e$ is the proton charge, $T_{s}$ is the background temperature, $\mu$ is the magnetic moment, and $\oint d \varphi$ signifies the gyroaverage at fixed particle position.

By running the code GS2 \cite{DorlandETGturb2000} on supercomputers, we can solve the gyrokinetic equation in a flux tube, a long narrow simulation domain that follows a single magnetic field line on a single flux surface of interest. This allows us to calculate the local radial fluxes of energy and toroidal angular momentum (per unit area) according to
\begin{align}
   Q_{s} &\equiv \left\langle \left\langle \left\langle \int d^{3} v \left( \frac{m_{s} v^{2}}{2} \right) \vec{v}_{E} ~ \delta f_{s} \cdot \frac{\Nabla \psi}{\left| \Nabla \psi \right|} \right\rangle_{\Delta} \right\rangle_{\psi} \right\rangle_{t} \\
   \Pi_{\zeta} &\equiv \left\langle \left\langle \left\langle \int d^{3} v \left( m_{i} R^{2} \vec{v} \cdot \Nabla \zeta \right) \vec{v}_{E} ~ \delta f_{i} \cdot \frac{\Nabla \psi}{\left| \Nabla \psi \right|} \right\rangle_{\Delta} \right\rangle_{\psi} \right\rangle_{t}
\end{align}
respectively. Here $m_{s}$ is the particle mass, $v$ is the speed, $\left\langle \ldots \right\rangle_{\Delta}$ indicates a coarse-grain average over a spatial scale that is larger than the turbulence and smaller than the device, and $\left\langle \ldots \right\rangle_{t}$ indicates a time average over the turbulent timescale. The flux surface average is defined by $\left\langle \ldots \right\rangle_{\psi} \equiv \int_{\psi} dS \left( \ldots \right)/\int_{\psi} dS$, where $S$ is the entire flux surface. However, to evaluate this in a local code, we restrict the integral to just the domain of our flux tube. This is acceptable because the flux tube (which runs from $0$ to $2 \pi$ in poloidal angle) can be extrapolated to fill in the entire surface (due to axisymmetry).


In general, to calculate the rotation profile in statistical steady-state one must invert
\begin{align}
   \Pi_{\zeta} \left( \Omega_{\zeta}, \frac{d \Omega_{\zeta}}{d \psi} \right) = 0 \label{eq:momFluxCond}
\end{align}
on every flux surface \cite{ParraIntrinsicRotTheory2015}. However, we can calculate a simple estimate of the level of rotation using local values of $\Pi_{\zeta}$ and $Q_{i}$ \cite{BallShafranovShift2016}. First, we assume that the energy flux is dominated by the diffusion of a temperature gradient \cite{FreidbergFusionEnergy2007pg452},
\begin{align}
   Q_{i} = -D_{Q} n_{i} \frac{d T_{i}}{d \psi} .   
\end{align}
Here we have neglected the drive from the density gradient, which should be acceptable given that we are focusing on toroidal ITG turbulence. Next, expanding equation \refEq{eq:momFluxCond} around $\Omega_{\zeta} = d \Omega_{\zeta} / d \psi =0$ and neglecting the term proportional to $\Omega_{\zeta}$ (which can only enhance the rotation \cite{PeetersMomPinch2007}) gives
\begin{align}
   \Pi_{\zeta} \left( 0, 0 \right) = D_{\Pi} n_{i} m_{i} R_{0}^{2} \frac{d \Omega_{\zeta}}{d \psi} .
\end{align}
Here $n_{i}$ is the ion number density, while $D_{Q}$ and $D_{\Pi}$ are the energy and momentum diffusion coefficients respectively. We can couple the energy and momentum transport by assuming a constant turbulent Prandtl number $Pr \equiv D_{\Pi} / D_{Q} \approx 0.7$, which is motivated by the results of previous numerical simulations \cite{BallMomUpDownAsym2014,BarnesFlowShear2011,HighcockRotationBifurcation2011}. This produces the estimate
\begin{align}
\frac{R_{0}}{v_{th, i}} \frac{d \Omega_{\zeta}}{d \psi} \approx \frac{-1}{2 Pr} \left( \frac{v_{th, i}}{R_{0}} \frac{\Pi_{\zeta}}{Q_{i}} \right) \frac{d}{d \psi} \Ln{T_{i}} , \label{eq:rotationGradEst}
\end{align}
where $v_{th, i}$ is the ion thermal velocity. This equation highlights the importance of the parameter $\left( v_{th, i} / R_{0} \right) \Pi_{\zeta} / Q_{i}$, which previous results indicate is fairly independent of the temperature gradient \cite{BallMomUpDownAsym2014,BallMomFluxScaling2016}. We see that, by finding the shape that maximizes $\left( v_{th, i} / R_{0} \right) \Pi_{\zeta} / Q_{i}$ at $\Omega_{\zeta} = d \Omega_{\zeta} / d \psi =0$, we can maximize the intrinsic rotation.

\section{Numerical results}
\label{sec:numResults}

To investigate the momentum transport in flux surfaces with both elongation and triangularity, we will first use the more realistic geometry of equation \refEq{eq:fluxSurfSpecGlobal} and perform a two dimensional scan in $\theta_{2}$ and $\theta_{3}$. We choose to fix $C_{N 2} = 0.45$ and $C_{N 3} = 0.1$ because these values approximate the ITER boundary shape when $\theta_{2} = 0$ and $\theta_{3} = \pi / 3$. By setting $\rho = 0.75$, we select a flux surface that is in the core, but is close enough to the boundary to have substantial triangularity (specifically it selects the third flux surface from the center in Figs. \ref{fig:cutawayTokamak} and \ref{fig:coilCurrents}). The major radius of $R_{0} = 3$ and minor radius of $a = 1$ are selected to give the tokamak a conventional aspect ratio. The safety factor of $q = 1.4$, magnetic shear of $\hat{s} \equiv \left( \rho / q \right) d q / d \rho = 0.8$, and background density gradient of $d \Ln{n_{s}} / d \rho = - 0.733$ are taken from the widely-used Cyclone base case \cite{DimitsCycloneBaseCase2000}. Note that setting this value of the magnetic shear is formally inconsistent with the constant current assumption used to derive our flux surface specification. However, all geometrical quantities appearing in gyrokinetics are calculated using the Miller local equilibrium model \cite{MillerGeometry1998}, which minimizes this error in the vicinity of our flux surface of interest.

Because the flux surfaces are so strongly shaped, we will use a large background temperature gradient of $d \Ln{T_{s}} / d \rho = - 3.0$ to ensure that the turbulence is driven unstable. For these values of the gradients, the dominant drive of turbulence is the ion temperature gradient (ITG). Additionally, these simulations treat the electrons as gyrokinetic, so the effects of the electron temperature gradient (ETG) and trapped electron modes (TEM) are included. However, no attempt was made to study these sub-dominant modes. That being said, we don't expect ETG to drive rotation because the ions (which carry most of the momentum of the plasma) behave adiabatically at the electron gyroradius scale. Additionally, we believe that the momentum driven by TEM modes should behave similarly to ITG (i.e. have the same sign and similar dependences). This is because (as we will soon see) the dominant effect driving rotation appears to be the interaction between shaping and toroidicity. The effect of toroidicity, that the modes peak on the outboard side of the device, is the same for ITG and TEM modes. However, this is speculative and more work is needed to explore this.

Because of the strong shaping, these simulations are computationally expensive. Properly resolving the sharp features of the flux surface required a fine grid along the magnetic field line. The simulations used 128 grid points, a factor of four larger than conventional. The radial wavenumber grid had 127 points and varied from $k_{r} \rho_{i} \in \left[ -2.52, 2.52 \right]$ in steps of $\Delta k_{r} \rho_{i} = 0.04$. Similarly, the poloidal wavenumber grid (which parameterizes the direction perpendicular to the field line, but within the flux surface) had 22 points and varied from $k_{p} \rho_{i} \in \left[ 0, 0.84 \right]$ in steps of $\Delta k_{p} \rho_{i} = 0.04$. These grid spacings were chosen using a convergence study, though the poloidal grid spacing is somewhat larger than usual (corresponding to seven ion gyroradii). To double-check, our results were verified by rerunning the code with a finer poloidal grid for two of the geometries. The remaining coordinates used more typical resolutions. In velocity space, the energy grid had 12 points and the untrapped pitch angle grid had 20 points. The trapped pitch angle resolution was determined by making each point in the poloidal grid a bounce point for particles traveling in both directions along the magnetic field line. Therefore, near the location of the maximum magnetic field there are only 2 trapped pitch angles, while near the location of the minimum magnetic field there are 129 trapped pitch angles.

\begin{figure}
	\centering
	\vspace{0pt}
	(a) \raisebox{-\height}{\includegraphics[width=0.9\textwidth]{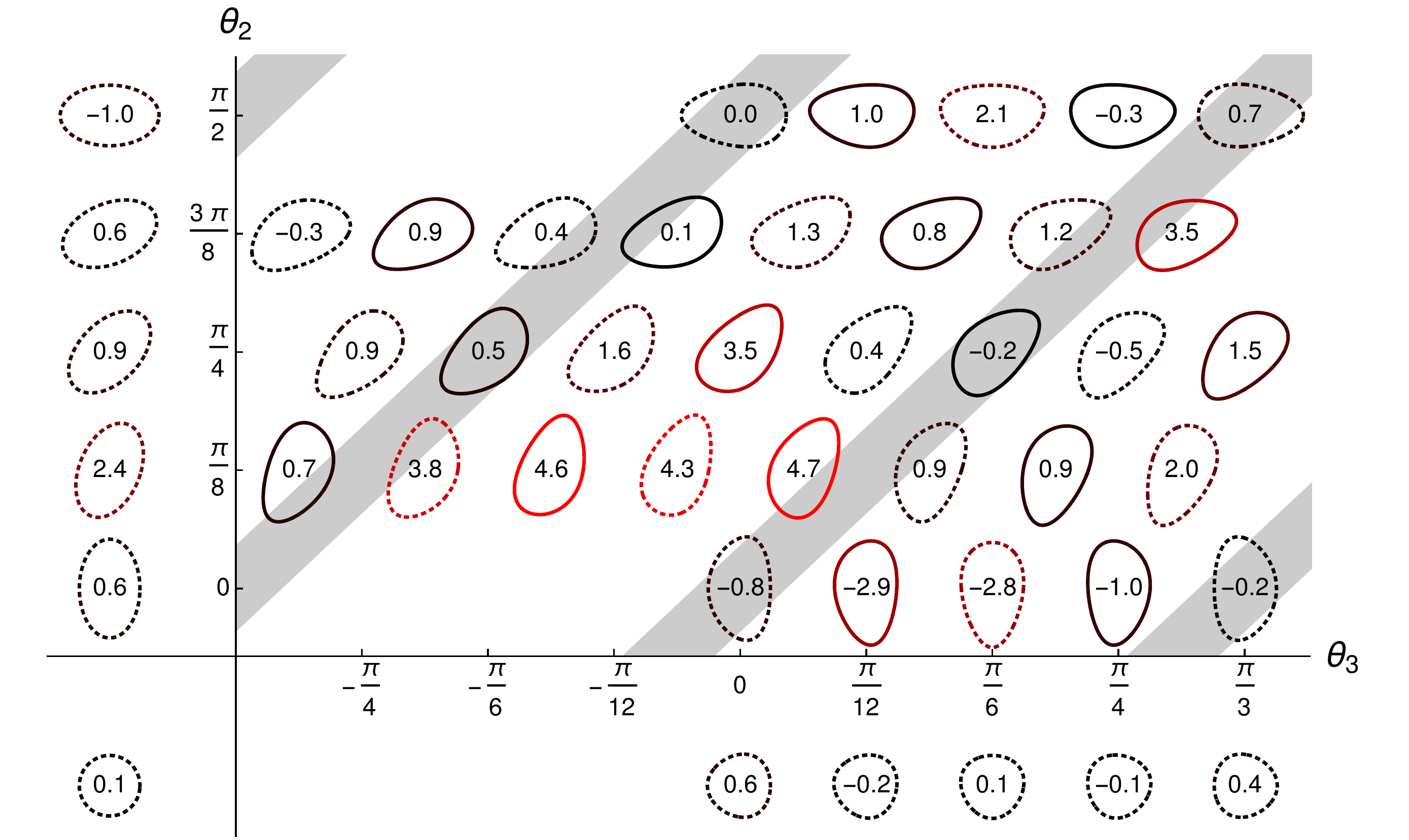}}
	
	(b) \raisebox{-\height}{\includegraphics[width=0.9\textwidth]{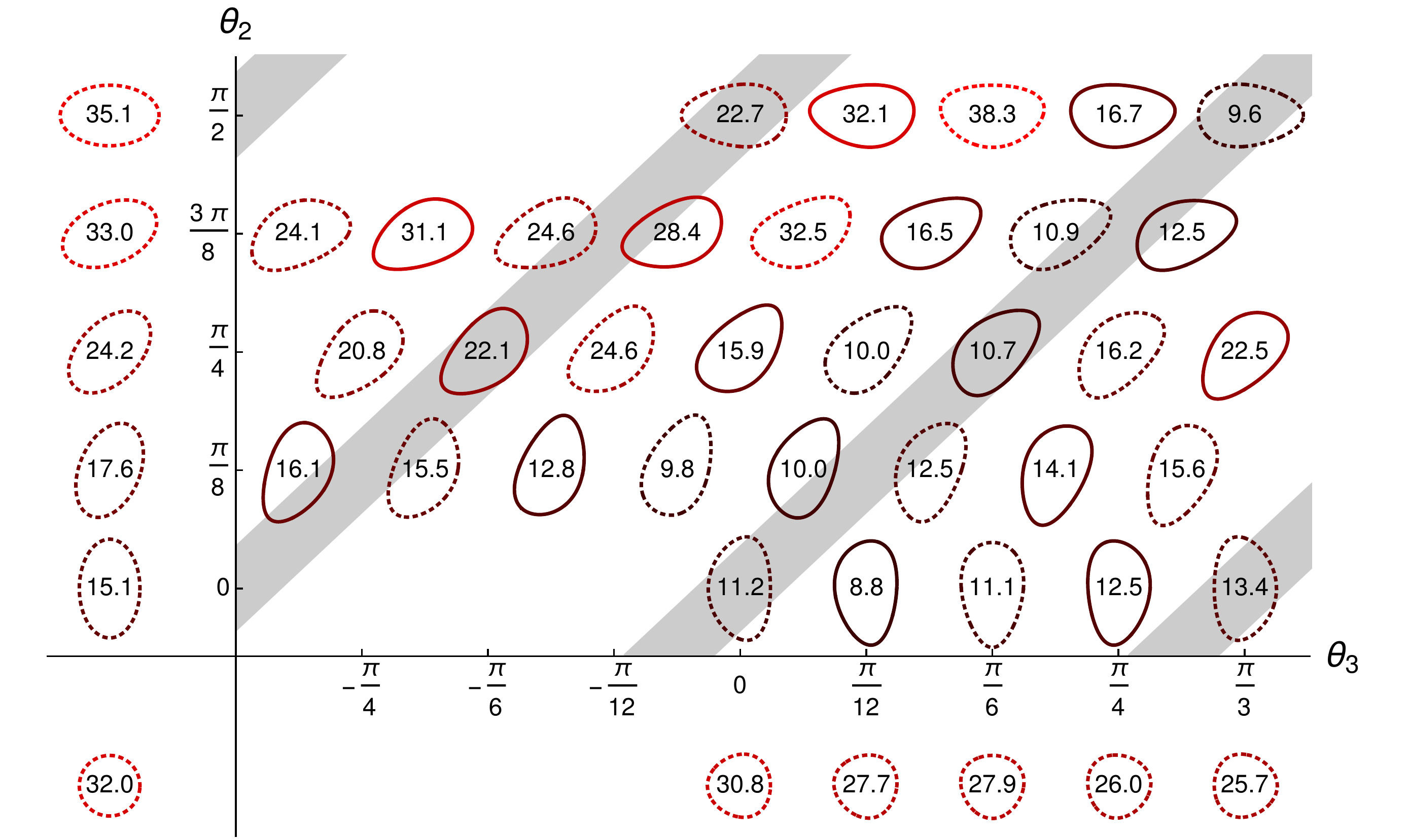}}
	\caption{Values of (a) the normalized momentum transport $100 \times \left( v_{th, i} / R_{0} \right) \Pi_{\zeta} / Q_{i}$ and (b) the total energy flux $\left( Q_{i} + Q_{e} \right) / Q_{gB}$ are indicated by the numbers/colors for various non-mirror symmetric (solid lines) and mirror symmetric (dotted lines) flux surfaces created with elongation and triangularity (specified using equation \refEq{eq:fluxSurfSpecGlobal}). The thick gray bands indicate geometries with up-down symmetric envelopes. For comparison, purely elongated flux surfaces are shown in quadrant \romanNum{2}, a circular flux surface is shown in quadrant \romanNum{3}, and purely triangular flux surfaces are shown in quadrant \romanNum{4}. Repeating several of these simulations indicates that the statistical error from averaging over the turbulent timescale is $\pm 0.5$ for the energy flux and $\pm 1$ for the normalized momentum transport (which is consistent with the small, but non-zero results of the up-down symmetric geometries).}
	\label{fig:nonMirrorGlobal}
\end{figure}

Figure \ref{fig:nonMirrorGlobal}(a) shows how the intrinsic rotation generated by the two-mode geometries compares with that generated by flux surfaces with only elongation or triangularity. We see that configurations shaped by only elongation produce significant momentum transport, unlike configurations with only triangularity. However, the ``optimal'' two-mode geometry (i.e. the $\theta_{2} = \pi / 8$, $\theta_{3} = \pi / 24$ case) has almost double the momentum transport of any single-mode geometry. Performing an additional simulation of the ``optimal'' geometry with $\left( R_{0} / v_{th, i}\right) d \Omega_{\zeta} / d \rho = 0.1$ confirmed that $\Pi_{\zeta} \approx 0$ as predicted by equation \refEq{eq:rotationGradEst}. This gives confidence that the assumptions used in our derivation (i.e. diffusive transport and the invariance of the Prandtl number) are well satisfied. If we assume that $\left( v_{th, i} / R_{0} \right) \Pi_{\zeta} / Q_{i}$ is uniform across the flux surfaces that have a substantial temperature gradient, then we can integrate equation \refEq{eq:rotationGradEst}. Without any edge rotation, this estimate predicts that the on-axis intrinsic rotation in the ``optimal'' geometry will have an $M_{S} \approx 7\%$. This value corresponds to $M_{A} \approx 1.2\%$, given an ITER-like value of $\beta = 0.06$ (i.e. the ratio of the thermal and magnetic pressures). This level of rotation is roughly what is needed to stabilize MHD modes.

One possible explanation for the large difference between the elliptical and two-mode geometries is the breaking of flux surface mirror symmetry. We know that breaking mirror symmetry allows the interaction of different shaping effects to directly drive momentum (which would be the dominant mechanism in a cylindrical device). However, for the aspect ratio of these simulations, this effect appears to be fairly modest. Of the four configurations with the most momentum transport, two of them are mirror symmetric.

Another possible explanation is that the beating between elongation and triangularity creates an $m=1$ mode (i.e. an envelope), which then interacts with toroidicity to drive the extra rotation. This would be the dominant momentum transport mechanism in configurations that only have high $m$ shaping effects. However, for the $m = 2$ and $m = 3$ mode numbers used here, this effect appears to be small. We can see that, of the four configurations with the most momentum transport, two of them have envelopes that are very close to up-down symmetric. This result is intuitive because it is difficult to visually discern any sort of envelope in the flux surface shapes. This is because the envelope is only distinct from the shaping effects when the difference between the two beating modes is much smaller than the mode numbers themselves. Instead, it appears that the intrinsic rotation drive is dominated by the direct interaction of elongation and triangularity with toroidicity.

Figure \ref{fig:nonMirrorGlobal}(b) shows that the best performing two-mode geometries stabilize turbulence and increase the confinement time. Because these simulations were all run at $\Omega_{\zeta} = d \Omega_{\zeta} / d \psi =0$, figure \ref{fig:nonMirrorGlobal}(b) shows only the direct effect of the flux surface shape. It does {\it not} include any beneficial effects that high levels of rotation might have on the turbulence. Even so, the ``optimal'' geometry has $25 \%$ less energy transport than the ITER-like shape (i.e.  $\theta_{2} = 0$ and $\theta_{3} = \pi / 3$). Furthermore, the turbulence in the geometries with both elongation and triangularity was completely stabilized at the Cyclone base case temperature gradient of $a / L_{T s} = 2.3$, unlike in the circular geometry. This demonstrates that the critical gradient was increased substantially by the strong shaping. Additionally, looking at the effect of positive and negative triangularity, we see behavior that is roughly consistent with TCV results \cite{WeisenShapeOnConfinement1997,MarinoniTCVtri2009}. Elongated configurations see more of a benefit from negative triangularity. Rigorous agreement would not be expected because of the importance of TEM turbulence and electron transport in TCV. In our simulations, the electron energy flux was consistently smaller than the ion energy flux, typically by a factor of four.

\begin{figure}
	\centering
	\vspace{0pt}
	(a) \raisebox{-\height}{\includegraphics[width=0.95\textwidth]{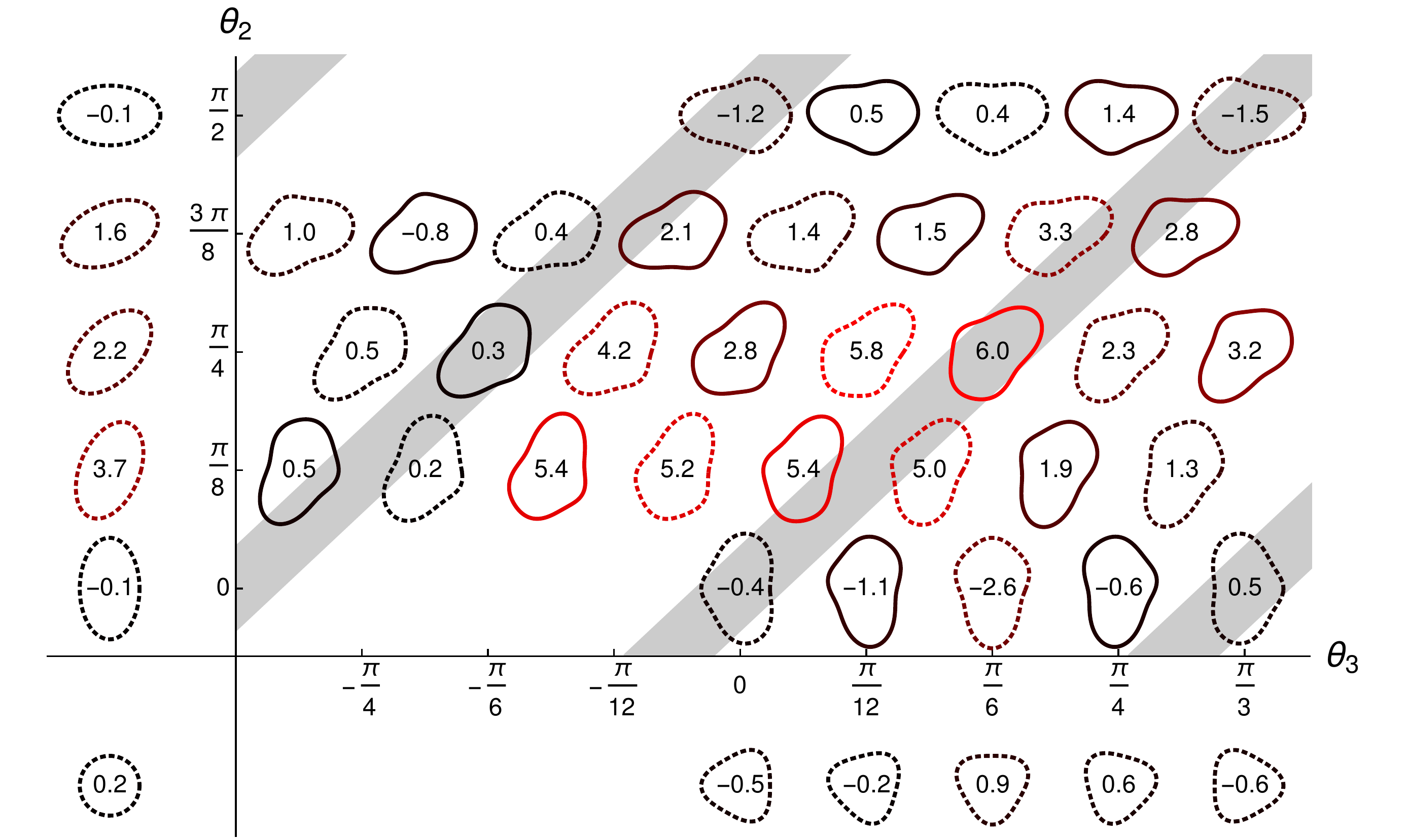}}
	
	(b) \raisebox{-\height}{\includegraphics[width=0.95\textwidth]{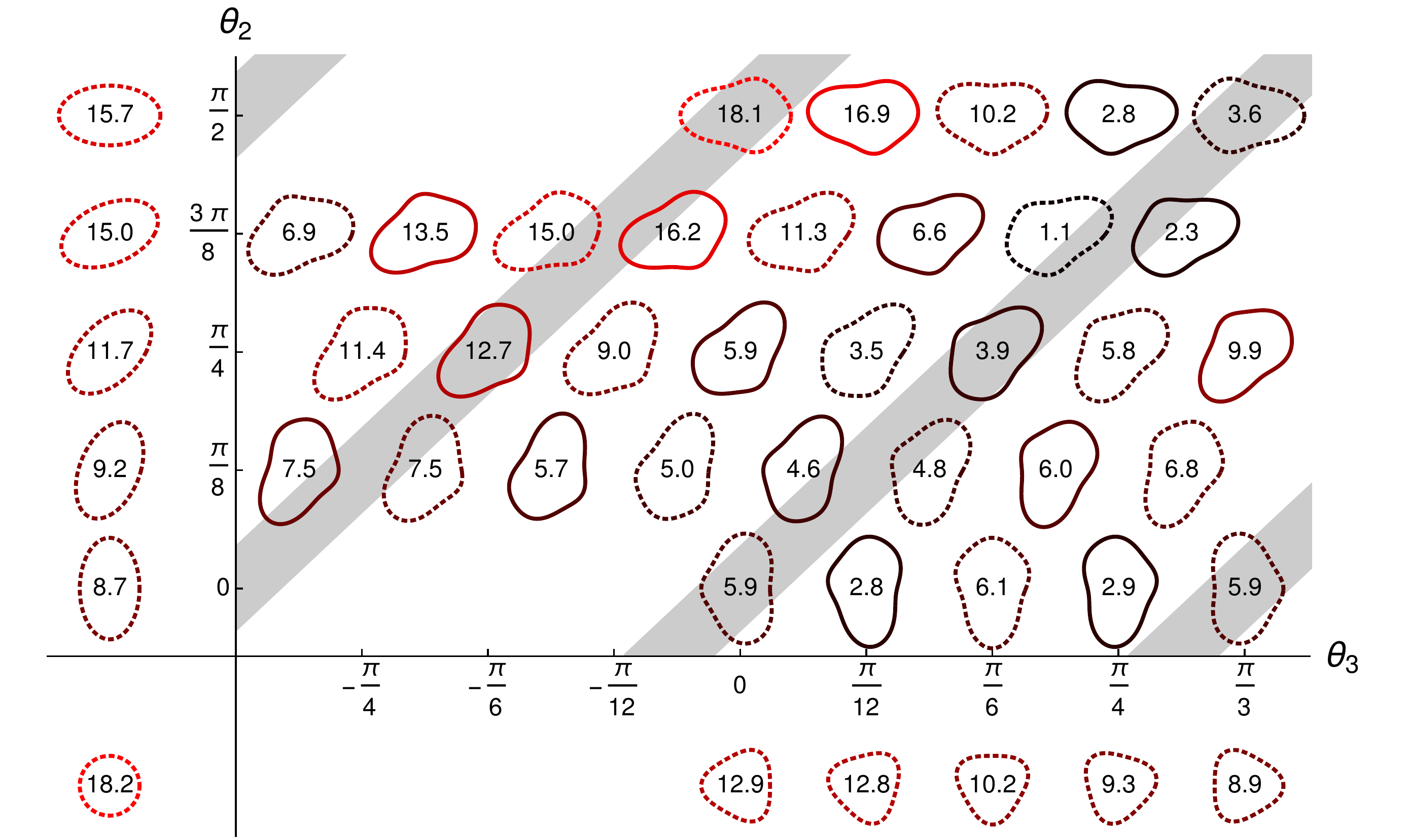}}
	\caption{Values of (a) the normalized momentum transport $100 \times \left( v_{th, i} / R_{0} \right) \Pi_{\zeta} / Q_{i}$ and (b) the total energy flux $\left( Q_{i} + Q_{e} \right) / Q_{gB}$ are indicated by the numbers/colors for various non-mirror symmetric (solid lines) and mirror symmetric (dotted lines) flux surfaces created with elongation and triangularity (specified using equation \refEq{eq:fluxSurfSpecGenElong}). The thick gray bands indicate geometries with up-down symmetric envelopes. For comparison, purely elongated flux surfaces are shown in quadrant \romanNum{2}, a circular flux surface is shown in quadrant \romanNum{3}, and purely triangular flux surfaces are shown in quadrant \romanNum{4}. The error is assumed to be similar to the error in figure \ref{fig:nonMirrorGlobal}.}
	\label{fig:nonMirrorGenElong}
\end{figure}

Lastly, to assess the sensitivity of these numerical results, we repeated the scan using the shape parameterization given by equation \refEq{eq:fluxSurfSpecGenElong} with somewhat different parameters. We increased the aspect ratio of the flux surface of interest by setting major radius to $R_{0} = 3$ and $\rho = 0.54$. This was done because reference \cite{BallShafranovShift2016} indicates that the momentum transport is sensitive to the aspect ratio (specifically, it increases with aspect ratio). Additionally, the magnitude of the shaping was increased somewhat to $C_{N 2} = 0.5$ and $C_{N 3} = 0.4$. The results are shown in figure \ref{fig:nonMirrorGenElong}. Because of the differences in shape and aspect ratio, it is not particularly illuminating to compare the exact values between corresponding configurations. However, figure \ref{fig:nonMirrorGenElong} supports all of the conclusions we arrived at from figure \ref{fig:nonMirrorGlobal}. First, pure elongation, unlike pure triangularity, can drive significant momentum transport. Second, adding some triangularity to elongation (at certain tilt angles) can significantly enhance the momentum flux. Moreover, the tilt angles of elongation and triangularity for the best-performing geometries are similar. Third, the effect of non-mirror symmetry and up-down asymmetric envelopes appear to be small. Fourth, many of the tilted configurations display a reduction in energy transport compared to the ITER-like shape (i.e.  $\theta_{2} = 0$ and $\theta_{3} = \pi / 3$). We note that in the upper-right region of figure \ref{fig:nonMirrorGenElong}(b), there are several configurations with particularly low energy transport. This feature can also be seen in figure \ref{fig:nonMirrorGlobal}(b) (though it is not quite as dramatic).

\section{Conclusions}
\label{sec:conclusions}

This work indicates that up-down asymmetry can drive sufficient intrinsic rotation to stabilize MHD modes in large devices. Our analysis has identified the optimal tilt angles (i.e. $\theta_{2} = \pi / 8$ and $\theta_{3} = \pi / 24$) to maximize the rotation for typical values of elongation and triangularity. Furthermore, we have shown that experimental coil sets can create this shape and, for the parameters used, it has 25\% less turbulent energy transport than a conventional ITER-like shape.

\ack

The authors would like to thank S. Brunner, I. Pusztai, M. Wensing, H. Reimerdes and P. Helander for useful discussions pertaining to this work. This work was funded in part by the RCUK Energy Programme (grant number EP/I501045). It has been carried out within the framework of the EUROfusion Consortium and has received funding from the Euratom research and training program 2014---2018 under Grant Agreement No. 633053. Computing time for this work was provided by the Helios supercomputer at IFERC-CSC under the projects SPIN, TRIN, and GKMSC. Additionally, we acknowledge the CINECA award under the ISCRA initiative, for the availability of high performance computing resources and support.

\appendix

\section*{References}
\bibliographystyle{unsrt}
\bibliography{references_OLD.bib}

\end{document}